\documentclass[twocolumn,floatfix,prx,superscriptaddress]{revtex4}

\usepackage{color}
\usepackage{array}
\usepackage{amsmath}
\usepackage{amssymb}
\usepackage{wasysym}
\usepackage{bm}
\usepackage{graphicx}
\usepackage{txfonts}
\usepackage{epsfig}
\usepackage{extarrows}

\newcommand{\abs}[1]{|#1|}
\newcommand{\bra}[1]{\langle \, #1 \,|}
\newcommand{\ket}[1]{|\, #1 \, \rangle}

\newcommand{\ketbra}[1]{\ket{#1} \bra{#1}}

\newcommand{\dg}{\dag}

\newcommand{\be}{\begin{equation}}
\newcommand{\ee}{\end{equation}}
\newcommand{\bc}{\begin{center}}
\newcommand{\ec}{\end{center}}

\newcommand{\non}{\nonumber}

\begin{document}
\title{Photon pair condensation by engineered dissipation}
\author{Ze-Pei Cian}
\altaffiliation{Contributed equally}
\affiliation{Department of Physics,University of Maryland, College Park, Maryland 20742, USA}
\affiliation{Joint Quantum Institute, NIST/University of Maryland, College Park, Maryland 20742, USA}
\author{Guanyu Zhu  \footnote{Email address: guanyuzhu2014@gmail.com. Current affiliation:  IBM T.J. Watson Research Center, Yorktown Heights, New York 10598, USA}}
\altaffiliation{Contributed equally}
\affiliation{Joint Quantum Institute, NIST/University of Maryland, College Park, Maryland 20742, USA}
\author{Su-Kuan Chu}
\affiliation{Department of Physics,University of Maryland, College Park, Maryland 20742, USA}
\affiliation{Joint Quantum Institute, NIST/University of Maryland, College Park, Maryland 20742, USA}
\affiliation{Joint Center for Quantum Information and Computer Science, NIST/University of Maryland, College Park, Maryland 20742, USA}
\author{Alireza Seif}
\affiliation{Department of Physics,University of Maryland, College Park, Maryland 20742, USA}
\affiliation{Joint Quantum Institute, NIST/University of Maryland, College Park, Maryland 20742, USA}
\author{ Wade DeGottardi}
\affiliation{Joint Quantum Institute, NIST/University of Maryland, College Park, Maryland 20742, USA}
\affiliation{Institute for Research in Electronics and Applied Physics, University of Maryland, College Park, MD 20742, USA}
\author{Liang Jiang}
\affiliation{Departments of Applied Physics and Physics, Yale University, New Haven, Connecticut 06511, USA}
\affiliation{ Yale Quantum Institute, Yale University, New Haven, Connecticut 06511, USA}
\author{Mohammad Hafezi}
\affiliation{Joint Quantum Institute, NIST/University of Maryland, College Park, Maryland 20742, USA}
\affiliation{Department of Physics,University of Maryland, College Park, Maryland 20742, USA}
\affiliation{Institute for Research in Electronics and Applied Physics, University of Maryland, College Park, MD 20742, USA}

\begin{abstract}
  Dissipation can usually induce detrimental decoherence in a quantum system.  However, engineered dissipation can be used to prepare and stabilize coherent quantum many-body states.  Here, we show that by engineering dissipators containing photon pair operators, one can stabilize an exotic dark state, which is a condensate of photon pairs with a phase-nematic order. In this system, the usual superfluid order parameter, i.e. single-photon correlation, is absent, while the photon pair correlation exhibits long-range order.  Although the dark state is not unique due to multiple parity sectors, we devise an additional type of dissipators to stabilize the dark state in a particular parity sector via a diffusive annihilation process which obeys Glauber dynamics in an Ising model. Furthermore, we propose an implementation of these photon-pair dissipators in circuit-QED architecture.

\end{abstract}
\pacs{***}
\maketitle

With the rapid development of quantum optical technology and quantum information platforms such as cavity/circuit quantum electrodynamics (QED) \cite{Blais:2007hh, Schoelkopf:2008vi,  houck2012} and Rydberg polaritons \cite{Saffman:2010ky}, it is now possible to investigate strongly-correlated many-body physics of photons  \cite{houck2012, Hartmann:2008ei, carusotto2013quantum, noh2016quantum, cardenas2019parity}.   While photons can have strong interactions in these platforms, they do not naturally thermalize, and one has to synthesize thermalization and a chemical potential to obtain many-body ground states \cite{Hafezi:2015jn, Kapit:2014vba,  ma2019dissipatively, Schiro:2012hr, wang2019theory, wang2018photon}.   Remarkably,  dissipation induced by the environment, which is usually regarded as a noise source leading to decoherence of the states, can actually become a useful resource.  If harnessed properly, dissipation can be used to autonomously prepare and stabilize an exotic many-body pure state as the steady/dark state of a system \cite{plenio1999cavity, Diehl:2008ha, Kraus:2008jda, Verstraete:2009kc, Weimer:2010ez, Cho:2011gd, Kastoryano:2011hr, Diehl:2011eh, Barreiro:2012jq, Reiter:2013ev, Rao:2014ka, Budich:2015cla, bardyn2012majorana, stannigel2014constrained}.   In the context of analogue quantum simulation, some well-known examples of dissipative  engineering schemes include the autonomous preparation and stabilization of the Bose-Einstein condensate (BEC) state \cite{Diehl:2008ha}, the Majorana-fermion state \cite{Diehl:2011eh} and the Chern insulator state \cite{Budich:2015cla}, all of which can be thought as the ground states of non-interacting Hamiltonians.  In a digital quantum simulation scheme \cite{barends2015digital, barends2016digitized, salathe2015digital}, a class of jump operators has been realized \cite{Kraus:2008jdb, Weimer:2010ez, barreiro2011open},  where the steady states correspond to the ground states of a specific class of interacting Hamiltonians.

Meanwhile, there have been significant experimental achievements in engineering analogue dissipators with higher-order photon jumps in small circuit-QED systems \cite{Murch:2012er, Shankar:2014jt,  Mirrahimi:2014js, Leghtas:2015fk,  Mundhada:2017id, Albert2018, Touzard:2018dia}. While these efforts have been motivated by autonomous error correction for a single- or two-sites system, it is interesting to investigate engineering many-body states, using these tools. Specifically, one can ask whether a strongly-correlated pure many-body state can be stabilized with an engineered analogue dissipator?  In this Letter, we answer this question by proposing a type of two-photon jump operator which can dissipatively prepare and stabilize an exotic strongly correlated photon-pair condensate exhibiting phase-nematic order. Furthermore, we propose an analogue experimental realization with circuit-QED systems. The added benefit of this approach is that one does not require effective thermalization or generation of a chemical potential in a photonic system.

\begin{figure}[t]
 \includegraphics[width=1.\columnwidth]{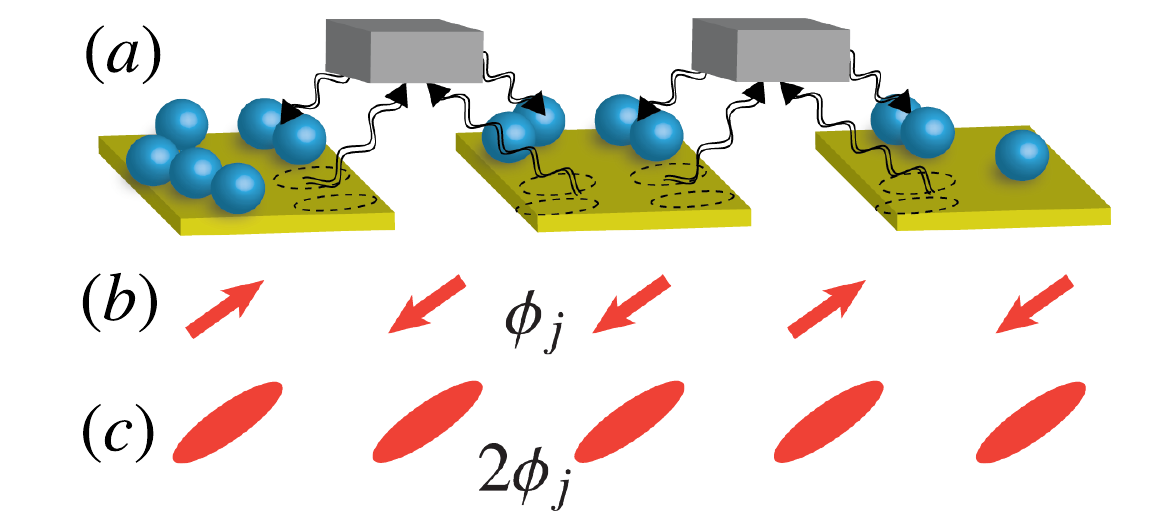}
  \caption{(a) Illustration of the dissipative process described by the photon pair jump operator.  (b) The U(1) phase variable $\phi_j$ (illustrated by the arrows) in the photon pair condensate is disordered due to the freedom to fluctuate by $\pi$.  (c) Twice of the U(1) phase angle $2\phi_j$ (illustrated by the rod) is ordered,  corresponding to a phase-nematic order.}
\label{fig:illustration}
\end{figure}
 
In order to illustrate the key idea, we start with a canonical example of Ref.~\cite{Diehl:2008ha}. For an open quantum system with Markovian environment, the system dynamics can be described by the Lindblad master equation:
\be\label{master_equation}
\frac{d}{dt}\rho  = -i [H, \rho] +  \mathcal{L}\rho,
\ee
where $H$ is the Hamiltonian of the system and the Liouvillian $\mathcal{L}\rho  = \sum_j  \kappa_j (2 l_j \rho l^\dag_j - l^\dag_j l_j \rho - \rho l^\dag_j l_j)$ describes the dissipation associated with the jump operator $l_j$ with decay rate $\kappa_j$. Consider the dynamics of bosonic particles on 1-dimensional lattice with $H = 0$ and a number-conserving jump operator of the form $l_{j}$$=$$(a^\dag_j + a^\dag_{j+1})(a_j - a_{j+1})$, connecting nearest neighbors in the lattice, where $a^\dag_j $ is the boson creation operator on site $j$.  These jump operators stabilize a BEC with fixed number of particles, i.e., a pure dark state  $\ket{D} =(a_{k=0}^\dag)^{N_{tot}} \ket{0} \propto \big(\sum_j a^\dag_j\big)^{N_{tot}} \ket{0}$, where $N_{tot}$ is the total number of bosons.  A simple way to understand these jump operators is the mean-field picture in which $a_j \rightarrow \sqrt{\bar{n}}e^{i \phi_j}$, where $\phi_j$ represents the compact U(1) phase variable (mod $2\pi$). 

The dark-state condition $l_j \ket{D}=0$ gives rise to the mean-field solution: $\phi_{j+1}-\phi_{j}= 0$ (mod $2\pi$), suggesting a phase locking between neighboring sites. 
   In this case, the dark state has long-range order, i.e., 
$\langle a^\dag_i a_j \rangle  $ $\xLongrightarrow{\abs{i-j} \rightarrow \infty} $ $  \langle a^\dag_i \rangle \langle a_j \rangle $ $ = \bar{n} $,
where we have introduced mean-field order parameter $\langle a_j \rangle$ $\approx$ $\sqrt{\bar{n}} \langle e^{i \phi} \rangle$ and $\phi$ is the uniform phase, after the spontaneous breaking of a $U(1)$ symmetry.  While this order parameter is fragile in 1D, these ideas can be generalized to higher dimensions where the long-range order can become robust.

\textit{Pair jump operators.}--- In this work, we propose a quartic jump operator connecting site $j$ and $j+1$ of the form
\be\label{jump2}
l_j = (a^{\dag 2}_j + a^{\dag 2}_{j+1})(a^2_j - a^2_{j+1}),
\ee
in a 1D lattice, as shown in ~Fig.~\ref{fig:illustration}(a). This can be generalized to 2D and 3D by assigning a jump on each link of the lattice. 
  Before deriving the exact form of the wave function, we consider a mean-field solution in which we take $a_j\rightarrow \sqrt{\bar{n}}e^{i \phi_j}$. The dark-state condition $l_j \ket{D}=0$ hence gives rise to the mean-field solution: $2(\phi_{j+1}-\phi_{j})= 0$ (mod $2\pi$), suggesting a locking of twice the phase variables between neighboring sites. This leads to the mean-field order parameter  $\langle a_j^2 \rangle = \bar{n} \langle e^{2i \phi_j} \rangle$ rather than $\langle a_j \rangle$, as in the previous case. In fact, we now have $\langle a_j \rangle=0$.   
%Note that the phase $\phi_j$ is now defined mod $\pi$ instead of mod $2\pi$ in the previous case,  meaning that the $U(1)$ symmetry is broken down to a residual $\mathbb{Z}_2$ symmetry.   
For correlation functions, we get
\begin{align}\label{correlator}
\nonumber & \langle a^{\dag 2}_i a_j^2 \rangle \xLongrightarrow{\abs{i-j} \rightarrow \infty} \langle a^{\dag 2}_i \rangle \langle a_j^2\rangle = \bar{n}^2 \langle e^{i 2(\phi_i - \phi_j)} \rangle = \bar{n}^2, \\
& \langle a^\dag_i a_j \rangle=\bar{n} \langle e^{i (\phi_i - \phi_j)} \rangle=\bar{n} \langle e^{i (\phi_i - \phi_j)+\pi} \rangle=-\langle a^\dag_i a_j \rangle = 0,
\end{align}
for $ \ i \neq j$. While $(2\phi_j)$ exhibits long-range order, $\phi_j$ does not since $\phi_j$ can flip by $\pi$ and still satisfy the dark-state condition [c.f.~Fig.~\ref{fig:illustration}(b,c)]. This photon pair condensate exhibits phase-nematic order. A similar state has been studied in the context of Josephson junction arrays \cite{Doucot:2002et},the symmetry breaking phase of a photon pair hopping Hamiltonian \cite{lebreuilly2019stabilizing} and fragmented many-body state in the ultra-cold atomic system \cite{mueller2006fragmentation, bader2009fragmented, fischer2015photonic}. The oriented rods without an arrow head in Fig.~\ref{fig:illustration}(c) represent the local order parameter $\langle e^{i 2\phi_j} \rangle$ for such a state, which does not differentiate the $\pi$-phase flip of $\phi_j$ and corresponds to the spontaneous breaking of a $U(1)/\mathbb{Z}_2$ symmetry.

\textit{Exact solutions.}---
 For the system where the Hamiltonian $H = 0$ and the jump operator is described in Eq.~\eqref{jump2}.The steady state density matrix is given by $\rho_{ss}=\ketbra{D}$ where $\ket{D}$ is annihilated by the jump operators in Eq.~\eqref{jump2} satisfying $l_j \ket{D}=0.$

We find that the dark state $\ket{D_{2n}}$ can be described as a condensate of $n$ two-photon bound states:
\begin{eqnarray}
\ket{D_{2n}} \propto A^{\dg n} \ket{0},
\label{exact_dark}
\end{eqnarray}
where $A^\dg = \sum_j a_j^{\dg 2}(n_j + 1)^{-1}$ \footnote{ See Supplemental Material for more details on the derivation dark state exact solution.} is the creation operator of quasi-particles related to photon pair bound state and $n_j = a^\dag_j a_j$ is the on-site number operator. Note that the extra normalization factor $(n_j+1)^{-1}$ in the definition creation operator $A^\dagger$ only affects the relative weights of different photon pair spatial configurations, but not the essence of the pair condensation. 

One can easily see that the single-particle correlator $\langle a^\dag_i a_j  \rangle$ (for $i\neq j$) vanishes because $a_i \ket{D_{2n}}$ and $a_j \ket{D_{2n}}$ have zero overlap since the photon occupation on site $i$ and $j$ becomes odd respectively. On the other hand, the pair correlation $\langle a^{\dag 2}_i a_j^2  \rangle$ is flat since $a_i^2 \ket{D_{2n}}$$=$$ a_j^2 \ket{D_{2n}}  \propto  \ket{D_{2(n-1)}}$ due to the fact that taking a pair out of the condensate at any site results in the same condensate with $n-1$ pairs of photons. This is just a manifestation of the definition of a pair condensate.   
 
\begin{figure}[t]
\includegraphics[width=0.90\columnwidth]{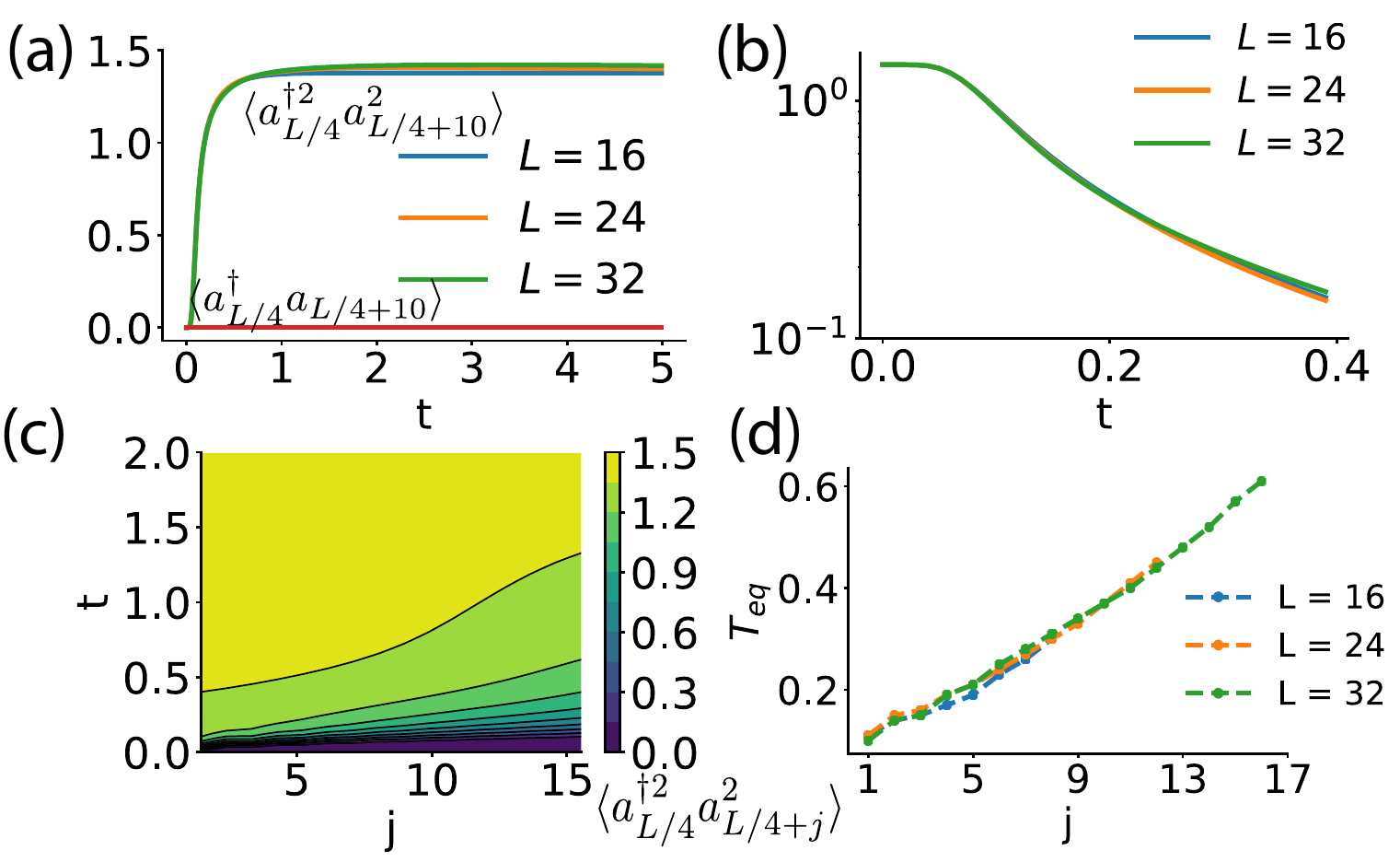}
    \caption{(a) Time evolution of single-photon and photon pair correlators as a function of time with various system size $L$.  The average photon density $\bar{n}=1$. (b) Time evolution of $\langle{a^\dg_{L/4}}^2(t$$=$$\infty) a_{L/4+10}^2(t$$=$$\infty)\rangle$$-$$ \langle {a^\dg_{L/4}}^2(t) {a^2_{L/4 + 10}}(t)\rangle$. Exponential saturation of the pair correlators (plotted in log scale) shows the dark-state cooling time is independent of system size. (c) Pair correlation function as a function of distance and time for $L = 32$ .  The unit of time is $\kappa^{-1}$. (d) Equilibrium time $T_{{\rm eq}}$ as function of distance $j$.  The linear dependence of the equilibrium time as function distance is due to the finite propagation speed of entanglement. }
\label{fig:figure_2}
\end{figure}

 We numerically simulate the time-dependent master equation Eq.~\eqref{master_equation} for an open 1D chain via quantum trajectory method with time-evolving block decimation (TEBD) algorithm \cite{gambetta_quantum_2008, Vidal:2004jc}, with results shown in Fig.~\ref{fig:figure_2}. We start with a product Fock state $\ket{2, 0, 2, 0, ...}$, and the jump operator drives the system to the steady (dark) state.  We see from Fig.~\ref{fig:figure_2}(a) that the single-particle correlation function $\langle a^{\dag }_{L/4} a_{L/4+10} \rangle$ remains zero at all times, while the pair correlation function $\langle a^{\dag 2}_{L/4} a_{L/4+10}^2  \rangle$ grows rapidly with an exponential saturation until reaching the steady state.   The whole time evolution resembles a cooling process. The cooling time  is independent of the system size, as seen in the plot where the total number of sites is varied as $L =16, 24, 32$.   The exponential saturation behavior and the cooling time is manifest on a logarithmic scale, see Fig.~\ref{fig:figure_2}(b).   
 
 We also plot the pair correlators as a function of the distance between two sites, i.e., $\langle a^{\dag 2}_{L/4} a_{L/4+j}^2  \rangle$ versus time $t$, as shown in Fig.~\ref{fig:figure_2}(c). We see that the state has almost flat correlation when reaching the dark state, consistent with the prediction from the analytical solution shown above.  Before reaching the dark state, the correlator is not flat and decays with distance. This is due to the fact that correlation between more distant sites needs more time to be built up. Fig.~\ref{fig:figure_2}(d) shows the equilibrium time $T_{eq}$ as function of distance. The equilibrium time $T_{{\rm eq}}$ is defined as the time it takes for the correlator $\langle a^{\dg 2}_{L/4} a^2_{L/4+j}  \rangle$ to reach $80\%$ of its steady state value. The spreading of the correlation function follows the Lieb-Robinson light cone beheavior.
 In addition, we have observed that the introduction of a Kerr non-linearity in the form $H=U a^{\dag 2}_j a_j^2$ in the system Hamiltonian leads to an exponential decay of correlator (in 1D) as a function of the distance $j-1$.  The decay becomes faster increasing $U$.

\textit{Parity sectors.}---The above analytical and numerical analyses only consider a simplified situation where the initial condition has all even number of photons. We note that even for fixed total photon number, the dark-state subspace has extensive degeneracies $2^{L-1}$. ($L$ is the number of sites), labeled by the local parity $P_j=(-1)^{n_j}$ on each site.  The exact wave function we wrote down above in Eq.~\eqref{exact_dark} is only the exact wave function for the sector where the parities of all sites are all even, i.e., $P_j=1$ for all $j$, which we call a ``pure pair condensate".  On top of that, there are odd-parity ``defects", which are created in pairs from the pure pair condensate.  

The wave function of a particular defect configuration can be given by $ \ket{D'_{n_d}} \propto \prod_{i=1}^{2n_d} a'^{\dag}_{d_i} \ket{D_{2(n-n_d)}}$, 
where $n_d$ denotes the number of pairs of odd-parity defects and their positions are labeled by $d_i$.  Several solutions of the parity-sector problem are discussed as follows. 

To begin with, we note that the different parity sectors are not coupled together via the jump operators.  Similar to what has been considered in the numerical simulation in Fig.~\ref{fig:figure_2}, one can start with an initial product state in the all-even sector (easy to prepare experimentally with pulses in the presence of onsite nonlinearity). In this case, the jump operator will only drive the system to the dark state $\ket{D_{2n}}$ in Eq.~\eqref{exact_dark}, in the absence of unwanted noise. We note that noise is always present in experimental systems which either brings the state to different sectors. Therefore, one only expects to prepare the targeting dark state with the jump operators before the unwanted decoherence dominates. If one aims to stabilize the dark state, extra measurement or stabilizing schemes are needed as discussed below.

 \begin{figure}[t]
\includegraphics[width=1\columnwidth]{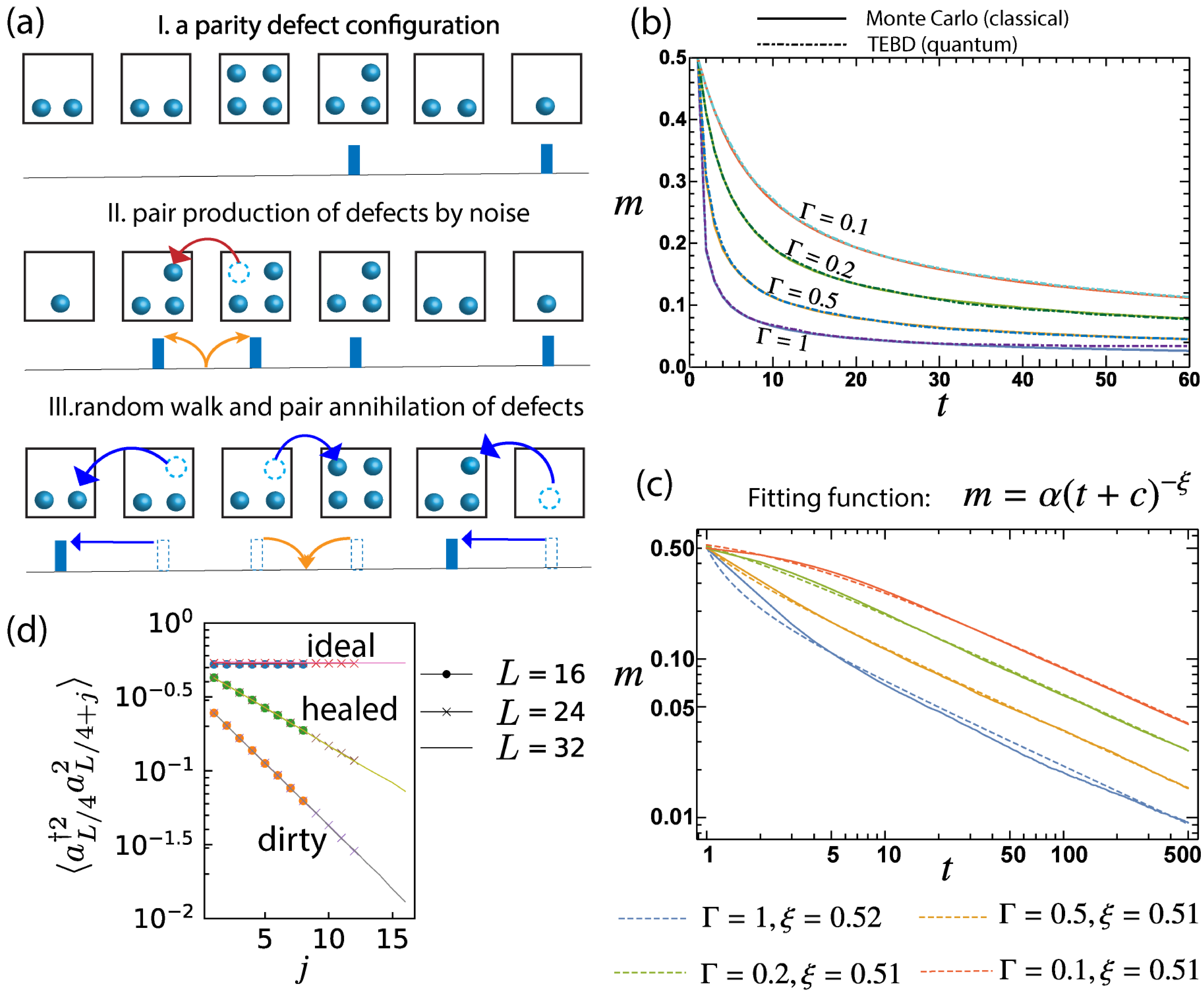}
 \caption{(a) Illustration: I.~particular odd parity defect (blue bars) configuration; II.~pair production of defects from single-photon hopping noise; III.~random walk of defects due to conditional hopping, and the induced pair annihilation process of defects (healing). (b)  Defect density ($m$) as a function of time ($t$). Classical Monte Carlo simulation (solid) with $L=100$ sites and quantum trajectory with matrix product state simulation (dot-dashed) with $L=30$ sites  of the time evolution of the average defect density. (c) Log-log plot of the classical Monte carlo simulation (solid) for $L=100$, with the fitted curves (dashed) showing the asymptotic power-law scaling $m(t)$ $\sim$ $ (\Gamma t)^{-1/2}$. (d) Pair-correlators normalized with the average photon density $\bar{n} = 0.5$ at different system size $L=16,24,32$ in the situations: \textbf{i}.~ideal (no noise); \textbf{ii}.~dirty case (with noise but no healing); \textbf{iii}.~healed case (with noise and healing). }
 \label{fig:figure_3}
 \end{figure}

A more general solution is: by imposing measurement and feedback operation on the parity of each site, i.e., $P_j=(-1)^{n_j}$, it is possible to keep projecting the many-body state to a particular parity sector.    A jump operator describing such measurement and feedback operation to stabilize parity-all-even sector is 
\be\label{parity_jump}
c_j = \Gamma (a^\dag_{j+1} a_{j} + a^\dag_{j-1} a_{j} )\left(\frac{1 - P_j}{2}\right) ,
\ee
where $\Gamma$ is the hopping rate. 
Note that this jump operator applies a hopping term connecting that particular site to its nearest neighbors conditioned by the the parity on the particular site being odd. It causes the odd-parity defect to take a random walk and eventually annihilate with another parity defect, as illustrated in Fig.~\ref{fig:figure_3}(a). This diffusive defect annihilation process resembles a chemical reaction described by the formula: $\mathbf{df}+\mathbf{df} \rightarrow 0$, where $\mathbf{df}$ stands for a single defect. Therefore, no defect will exist in the steady state if the total photon number is even ($2n$), and so the steady state becomes a pure pair condensate. We call such a process ``\textit{healing}".   

Since the parity measurement at time $t+dt$ will post-select the direction (left or right) to which the defect has hopped at time $t$, the defect dynamics can be exactly mapped  to a classical stochastic dynamics of the diffusive annihialation problem. For a 1D chain, the dynamics of defect density exhibits power law decay:  $m(t)$ $\sim$ =$(\Gamma t)^{-1/2}$ \cite{Lee:1994ic, Racz:1985fy} 

The classical Monte Carlo simulation (with 100 sites) of the time evolution of the average defect density $m(t)$ quantitatively agrees with the quantum TEBD simulation (with 30 sites) for a 1D chain, as shown in Fig.~\ref{fig:figure_3}(b). The former is plotted in a log-log scale in Fig.~\ref{fig:figure_3}(c), where the associted fitting curves confirms an asymptotic power-law   $m(t)$ $\sim (\Gamma t)^{-1/2}$.  We have checked in both types of simulations that the time evolution of $m(t)$ is almost independent of system size.

 In the presence of additional noise, such as incoherent single-photon hopping described by the jump operator $l'_j $$=$$ a^\dag_j a_{j+1}$, there exists a finite defect production rate $\Gamma h$. The defect production can be balanced by the diffusive defect annihilation process with hopping rate $\Gamma$, and the defect density approaches a residual steady-state density $m_s$ with a characteristic relaxation time $\tau$. We have the following scaling (when $h \ll 1$): $m_s \sim h^{1/\delta}$ and $\tau \sim \Gamma^{-1} h^{-\Delta}$ \cite{Racz:1985fy}. Note that even a large hopping rate $\Gamma$ cannot reduce the steady-state density $m_s$, but only the healing time $\tau$. A generic scaling law $\Delta\xi=1/\delta$ should be satisfied \cite{RACZ:1985kt}. For a 1D chain, we have $\delta =2$ and $\Delta=1$  \footnote{See Supplemental Material which includes  Ref.~\cite{Racz:1985fy, Glauber:1963ij, Felderhof:1971tj, RACZ:1985kt, Torney:1983cy, Lee:1994ic}}.
 
 In Fig.~\ref{fig:figure_3}(d), we use TEBD to calculate the steady-state pair correlators $\langle a^{\dag 2}_{L/4} a_{L/4+j}^2  \rangle$ for system sizes $L=16,24,32$, in the following situations: \textbf{i}.~ideal case (no noise: $h=0$); \textbf{ii}.~dirty case (in the presence of single photon hopping noise: $h>0$, without healing: $\Gamma=0$); \textbf{iii}.~healed case (with noise and healing: $h,\Gamma>0$).  We see that in the presence of noise which proliferate the parity defects, the steady state ends up with a mixed state and the correlators decays exponentially, while the healing process significantly slows down the decay.

Due to the complexity of the parity operator, one typically can only realize such a jump operator with an active parity measurement in circuit-QED setup through either continuous \cite{Cohen:2017ep} or discretized repeated  \cite{ofek2016extending, hu2018demonstration} measurement schemes, instead of using continuous autonomous stabilization.  Nevertheless, in the situation that we first impose hard-core condition for occupation more than three photons, i.e., $a^{\dag 3}_j=0$, the parity condition can be simply converted to occupation condition, and we can effectively re-express the jump operator in Eq.~\eqref{parity_jump} as $c'_j$$=$$(a^\dag_j a_{j+1}$$+$$\text{H.c.})n_j(n_j -2)$, up to a constant factor of 2.  This jump operator can potentially be implemented continuously and hence autonomously stabilize the targeting pure photon pair condensate.  Similarly, one can either actively or autonomously monitor and stabilize the total photon number in the system.

 \begin{figure}[t]
\includegraphics[width=0.4\textwidth]{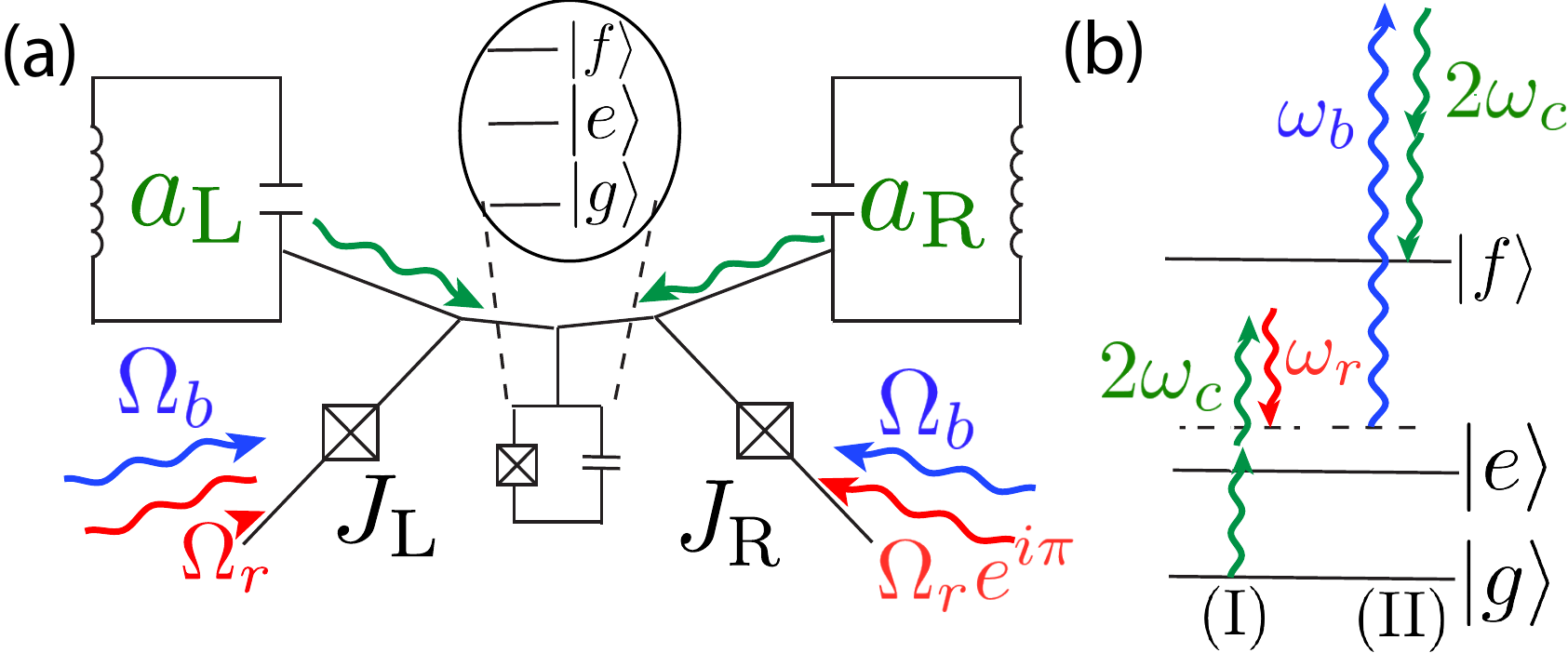}
 \caption{Proposed circuit-QED setup. (a) Two high-Q cavities at frequency $\omega_c$ and an three-level anharmonic oscillator($\ket{g}, \ket{e} $ and $\ket{f}$) coupled to the Josephson junction modes $J_L$ and $J_R$. Both of the Josephson junction modes are driven by a two-tone drive with frequency $\omega_r$ and $\omega_b$. (b) Process (I) shows a pairs of cavity photons combine with a driving $\omega_r$ to produce a virtual junction excitation $g \rightarrow e$. The effective coupling is given by $(a_L^2-a_R^2)\ket{e}\bra{g}$. The minus sign in the effective coupling comes from the relative phase of the Rabi frequency $\Omega_r$ as shown in (a). Process (II) gives the effective coupling $(a_L^{\dagger 2} +a_R^{\dagger 2})\ket{f}\bra{e}$. These two virtual processes are detuned from the respective resonance condition by a frequency $\delta$. The combination of these two processes yields the desired interaction. }
 \label{System}
 \end{figure}

Finally, we can also passively stabilize the parity sector via energetic constraint in the case of hard-core condition ${a^\dag}^3_j=0$. This is achieved by assign the following energy penalty term
$\delta H = -V_0 n_j (n_j - 2)$ in the Hamiltonian $H$ \cite{ Huang:2014dc}.
Therefore, the configuration with single-photon occupation (odd parity) on any site is projected out of the low-energy sector.

\textit{Experimental realization with Circuit QED.}---We illustrate the experimental scheme with a two-site jump operator $l$$=$$(a_L^{\dg 2} + a_R^{\dg 2})(a_L^{2}-a_R^{2})$. The generalization to a 1D chain is straightforward.  Consider a system consisting of the two high Q cavities $a_L$ and $a_R$, and an anharmonic oscillator.  The anharmonic oscillator is modeled by a three-level system ($\ket{g}, \ket{e}$ and $\ket{f}$) and is coupled to a cavity $a_\lambda$ and a Josephson junction mode $J_\lambda$, where $\lambda = L, R$, at both sides respectively. Both junction modes are driven by a two-tone drive $\Omega_\lambda(t) = \Omega_{\lambda,r}e^{i\omega_r t} + \Omega_{\lambda,b}e^{i\omega_b t}$ as shown in Fig. \ref{System}. 

We engineer a two-photon jump operator via four-wave mixing induced from the junction modes $J_L$ and $J_R$. The drive $\omega_r$ ($\omega_b$) is used to introduce an exchange of two photons of cavity mode $a^2_\lambda$ ($a^{\dg 2}_\lambda$) with the excitation $g \rightarrow e$ ($e \rightarrow f$). The four-wave mixing interaction of the pump $\omega_r$ ($\omega_b$) is proportional to $\sum_\lambda \Omega_{\lambda, r}a^2_\lambda \ket{e}\bra{g}$ ($\sum_\lambda \Omega_{\lambda, b}a^{\dg 2}_\lambda \ket{f}\bra{e}$). The minus sign in the jump operator can be engineered by introducing a $\pi$ phase shift between $\Omega_{L,r}$ and $\Omega_{R,r}$. The effective Hamiltonian is of the form
\begin{align}
H' = -\frac{\chi }{2}\sum_{\lambda = L,R}a_\lambda ^{\dagger 2}a_\lambda ^2 + g_1T_-\ket{e}\bra{g} + g_2T_+^\dg \ket{f}\bra{e} + h.c.
\notag
\end{align}
where $T_\pm = a^2_L \pm a^2_R$, $\chi$ is the Kerr nonlinearity induced from the junction modes and $g_1$ ($g_2$) is proportional to the Rabi frequency $\Omega_{r}$ ($\Omega_{b}$) as shown in Fig.~\ref{System}a. 

To obtain the jump operator in Eq.~\eqref{jump2}, we  combine the two-photon loss process ($T_-$) and two-photon creation process ($T_+^\dg$) in equation above. We can achieve this by detuning the two four-wave mixing processes by $\delta$ as shown in Fig.~\ref{System}b so that only a cascade of two such processes is possible \cite{Albert2018}. The detunning $\delta \gg g_1, g_2$ allows a two-photon exchange processes via a Raman transition. The effective Hamiltonian becomes 
\begin{align}
H_\text{eff} = -\frac{\chi }{2}\sum_{\lambda = L,R}a_\lambda ^{\dagger 2}a_\lambda ^2 + \frac{1}{\delta}
\begin{bmatrix}
|g_2|^2T_+^\dg T_+ & g_1 g_2 T_+^\dg T_- \\
g_1^* g_2^* T_-^\dg T_+ & |g_1|^2T_-^\dg T_-
\end{bmatrix}
\label{Heff}
\end{align}
where the 2 $\times$ 2 matrix acts on the anharmonic oscillator basis $\{ \ket{f}, \ket{g}\}$. In Eq. \eqref{Heff}, the term $T^\dg_+ T_- \ket{f}\bra{g}$ gives the desired two-photon process coupled to $g\leftrightarrow f$ transition. Assuming the decay rate($\kappa$) of process $f\rightarrow g$  is much greater than all of the other coupling constant in Eq. \eqref{Heff}. The system can be described 
$H_s =  -\frac{\chi }{2}\sum_{\lambda = L,R}a_\lambda ^{\dagger 2}a_\lambda ^2 + \frac{|g_1|^2}{\delta}(a_L^{\dagger 2}-a_R^{\dagger 2})(a_L^2-a_R^2)$ and the jump operator $l_s = \sqrt{\frac{2}{\kappa}}\frac{g_1 g_2}{\delta} (a_L^{\dagger 2}+a_R^{\dagger 2})(a_L^{2}-a_R^{2})$. The jump operator gives rise to Eq.~\eqref{master_equation} for the array case. The self-Kerr and the cross-Kerr nonlinearity terms in Hamiltonian $H_s$ can be eliminated by adding an extra pair of Josephson junction and a two-level system \footnote{ See Supplemental Material for more details.}.

 \textit{Conclusion and outlook.}--- We have discovered a photon pair jump operator which can dissipatively prepare and stabilize an exotic two-photon pair-condensate with phase-nematic order, with a circuit-QED implementation.  We have further proposed a conditional hopping operator to stabilize the dark state in a particular parity sector. Such a scheme can also be realized with Rydberg polaritons or ion-trap systems. An interesting future direction would be using such higher-order dissipators for autonomous quantum error correction using bosonic codes.   
 
\begin{acknowledgments}
 We thank Michel Devoret, and Rosario Fazio for insightful discussions. This work was supported by ARO-MURI, NSF-PFC at the JQI and the Sloan Foundation. Su-Kuan Chu is additionally supported by ARL-CDQI, NSF Ideas Lab on Quantum Computing, DoE BES Materials and Chemical Sciences Research for Quantum Information Science program, DoE ASCR Quantum Testbed Pathfinder program, and Studying Abroad Scholarship by Ministry of Education in Taiwan (R.O.C.). Liang Jiang acknowledge support from ARO, AFOSR, MURI and the Packard Foundation.
\end{acknowledgments}
 
 %%%%%%%%%%%%%%%%%%%%%%%%%%%%%%%%%%%%%%%%%%%%%
%\newpage

\begin{appendix}

\section{Exact Solution of the Dark State\label{append:AppDarkState}}
\subsection{Pure Two-Photon Condensate}
In this section, we are going to show that the analytical expression of the dark state of the jump operators $l_j = (a_{j}^{\dg 2} + a_{j+1}^{\dg 2})(a_{j}^2 - a_{j+1}^2)$, where $j$ is integer and $1 \leq j \leq L$, is 
\begin{align}
\ket{D_2m} = A^{\dg m}\ket{0} ,
\label{DarkState}
\end{align}
 where 
 \begin{align}
 A^\dg = \sum_{i = 1}^{L} a^{\dg 2}_i\frac{1}{n_i + 1},
 \end{align}
 and the number of photon is $2m$ . 
 
 We start by introducing two useful identities. (1) $[l_j, A^\dg] = P^1_{j} - P^1_{j+1}$. (2) $P^1_{j}A^\dg = 0$, where $l_j = a_{j} - a_{j+1}$ is the annihilation part of the jump operator $l_j$ and $P^1_{j}$ is the projector that project a state to one photon Fock state at site j, or $P^1_{j} = \ketbra{1_j}$. 
 
Proof of identity (1) : The commutator $[a^2_j, A^\dg]$ is
\begin{align}
[a^2_j, A^\dg] & = [a^2_j ,\sum_{i = 1}^{L}\sum_{n_i = 0}^\infty \sqrt{\frac{n_i+2}{n_i+1}}\ket{n_i + 2}\bra{n_i} ] \notag \\
& = \sum_{n_j = 0}^\infty (n_j + 2)\ket{n_j}\bra{n_j} - \sum_{n_j = 0}^\infty (n_j + 2)\ket{n_j+2}\bra{n_j+2} \notag \\
&= \sum_{n_j = 0}^\infty (n_j + 2)\ket{n_j}\bra{n_j} - \sum_{n_j = 2}^\infty n_j\ket{n_j}\bra{n_j} \notag \\
& = 2I + \ket{1_j}\bra{1_j} = 2I + P^1_j, 
\end{align}
where $I$ is the identity operator. 

Therefore, we have the commutator 
\begin{align}
[l_j, A^\dg] = [a^2_j - a^2_{j+1}, A^\dg] = P^1_j - P^1_{j+1}.
\end{align}

Proof of identity (2) : Since the $A^\dg$ operator always creates more than 1 photons, $A^\dg\ket{s}$ is always orthogonal to $\ket{1_j}$ for any arbitrary state $\ket{s}$ and hence $P^1_jA^\dg\ket{s}$ is always zero. Or equivalently, 
\begin{align}
 P^1_{j}A^\dg = \ket{1_j}\bra{1_j} \sum_{i = 1}^{L}\sum_{n_i = 0}^\infty \sqrt{\frac{n_i+2}{n_i+1}}\ket{n_i + 2}\bra{n_i} = 0.
 \end{align} 
 With these two useful identities, we can proceed to proof that Eq. (\eqref{DarkState}) is a dark state under jump operator $l_j$. 
 
Proof: When $m = 1$, we have
 \begin{align}
 l_j A^\dg\ket{0} = (A^\dg l_j +P^1_j - P^1_{j+1})\ket{0} = 0.
 \end{align}
 Here we use the identity (1). 
 
 Assume that for an integer $k$, $l_j (A^\dg)^{k}\ket{0} = 0$. 
 
 Then for $ m = k+1$, we have
 \begin{align}
 l_j  (A^\dg)^{k+1}\ket{0}  = (A^\dg l_j + P^1_j - P^1_{j+1}) A^{k}\ket{0} = 0.
 \end{align}
 The first term in the right hand side of above equation is zero because we assume $l_j (A^\dg)^{k}\ket{0} = 0$. The second and the third terms are zero because of the identity (2).
 
 By mathematical induction, equation $l_j (A^\dg)^m\ket{0} = 0$ holds for all integer $m$ greater than zero. And therefore, $\ket{D_2m}$ is the dark state under the dissipation $l_j$. 
 
\subsection{With Parity Defects}
 The dark state in the Eq \eqref{DarkState} is not unique in the absence of the healing process. One can always create localized parity defects as described in the main text. The dark state with parity defects of a particular defect configuration can be written as
 \begin{align}
\ket{D'_2m} = \prod_{i = 1}^{2n_d} a^{' \dg}_{d_i}  (A^\dg)^{n - n_d}\ket{0},
 \end{align}
 where $2n_d$ is the number of defect, $d_i$ is the position of the ith defect and 
 \begin{align}
a^{' \dg}_{d_i}  =a^\dg_{d_i} \frac{1}{n_{d_i} + 1}.
 \end{align}
 To show that this is also a dark state, let's consider the commutator $[l_i, a'_j]$. 
 
 If $i \neq j$ and $i \neq j-1$, then these two operators commute with each others. When $i = j$, we have
 \begin{align}
 [l_i, a'_j] & = [a^2_j, \sum_{n_j = 0}^\infty \frac{1}{\sqrt{n_j+1}} \ket{n_j+1}\bra{n_j}] \notag \\
 & = \sum_{n_j = 1}^\infty \sqrt{n_j} \ket{n_j-1}\bra{n_j} - \sum_{n' = 0}^\infty \sqrt{n'_j+2} \ket{n'_j+1}\bra{n'_j+2} \notag \\
 & = \ket{0_j}\bra{1_j}.
 \end{align}
Also, when $j = i+1$, the commutator becomes $ [l_i, a'_j]  = -  \ket{0_j}\bra{1_j}$. 

With this commutator, we can see that 
\begin{align}
l_j \ket{D'_2m} &= l_j \prod_{i = 1}^{2n_d} a^{' \dg}_{d_i}  (A^\dg)^{n - n_d}\ket{0} \notag \\
& = \prod_{i = 1}^{2n_d} [a^{' \dg}_{d_i}  l_j + \delta_{j,d_i} \ket{0_j}\bra{1_j} \notag\\
& - \delta_{j+1,d_i} \ket{0_{j+1}}\bra{1_{j+1}}] (A^\dg)^{n - n_d}\ket{0} \notag \\
&= 0.
\end{align}
The first term in the last equation above is zero because of the identity (1) and the second term and the third term vanishes since the state $(A^\dg)^{n - n_d}\ket{0} $ is always orthogonal to $\ket{1_j}$ by parity symmetry.

\section{Effective Classical Theory and Simulation for the Healing Process and the Steady States with Parity Defects}\label{append:stochastic}

In this section, we discuss the effective classical theory of the diffusive annihilation (healing) process of parity defects mentioned in the main text. 

 We start the discussion by focusing on the 1D situation, while the generalization to higher dimension is straightforward and will be discussed later.  We label the defect number on a 1D periodic lattice with $L$ sites by $\{\mathbf{m}\}$$\equiv$$\{m_1,...,m_j,m_{j+1},...,m_{L}\}$, where $j$ labels the sites.   For later convenience, we map the diffusive annihilation process onto a periodic classical Ising spin chain with $L$ sites as in Ref.~\cite{Racz:1985fy}.  The spin configuration is labeled by  $\{\sigma\}$$\equiv$$\{\sigma_1,...,\sigma_j,\sigma_{j+1},...,\sigma_{L}\}$, consisting of the stochastic spin variables $\sigma_j = \pm1$ ($1$ stands for  $\uparrow$ and $-1$ stands for $\downarrow$).  The correspondence between the spin and defect configuration is through the mapping $m_j=(1-\sigma_j \sigma_{j+1})/2$.   The presence of a defect at $j$ (i.e., $m_j=1$) is equivalent to the presence of a domain wall between opposite spins at site $j$ and $j+1$ on the spin chain, i.e.,  $\uparrow_j \downarrow_{j+1}$ or $\downarrow_j \uparrow_{j+1}$.  One can thus think the defect is living on the bond of the effective spin chain. The spin flip on the left/right of the domain wall corresponds to the moving of defect (domain wall) towards left/right, as illustrated by the processes $\uparrow\uparrow\downarrow\downarrow \Rightarrow \uparrow\downarrow\downarrow\downarrow$ and $\uparrow\uparrow\downarrow\downarrow \Rightarrow \uparrow\uparrow\uparrow\downarrow$ respectively.  The spin flip in the middle of two domain walls correspond to the annihilation of two defects (domain walls), as illustrated by the processes $\downarrow\uparrow\downarrow \Rightarrow \downarrow\downarrow\downarrow$ and  $\uparrow\downarrow\uparrow \Rightarrow \uparrow\uparrow\uparrow$.  The inverse of such annihilation processes gives rise to the pair production processes of defects (domain walls).

The probability distribution functions of the effective spin configuration are represented as $P(\{\sigma\},t)$.  There are $2^{L}$ such distribution functions in total, and they satisfy the following classical master equation first introduced by Glauber \cite{Glauber:1963ij}:
\begin{align}\label{master}
\non &\frac{d}{dt}P(\{\sigma\},t) =-\sum_j w_j(\{\sigma_{j-1},\sigma_j,\sigma_{j+1}\}) P(\{\sigma\},t) \\
&+\sum_j  w_j(\{\sigma_{j-1},-\sigma_j,\sigma_{j+1}\}) P(\{\sigma_1,...,-\sigma_j,\sigma_{j+1},...\},t).
\end{align}
Here, the first term on the right hand side describes the transition from the current configuration $\{\sigma_j\}$ at time $t$ to the new configuration $\{\sigma_1,...,-\sigma_j,\sigma_{j+1},...\}$ with the $j^\text{th}$ spin being flipped, while the second term describe the transition from the configuration $\{\sigma_1,...,-\sigma_j,\sigma_{j+1},...\}$ to the current configuration $\{\sigma_j\}$ by flipping the $j$th spin.  Therefore, all the stochastic dynamics are captured by a single spin flip.  The flipping rate ($w_j$) of the $j$th spin is determined by the local spin configuration involving the sites $j$ and $j\pm 1$. First, we require the defect to hop either to the left or right with hopping rate $\frac{1}{2}\Gamma$, leading to the following conditions:
\be\label{rate1}
w_j(\{\uparrow\uparrow\downarrow\})=w_j(\{\downarrow\downarrow\uparrow\})=w_j(\{\downarrow\uparrow\uparrow\})=w_j(\{\uparrow\downarrow\downarrow\})=\frac{1}{2}\Gamma.
\ee
Next, we require neighboring defects to annihilate with each other with rate $\Gamma$, which is given by the defect hopping rate and the fact that double occupation of the defects on the same site is equivalent to zero defect. This leads to  
\be\label{rate2}
w_j(\{\uparrow\downarrow\uparrow\})=w_j(\{\downarrow\uparrow\downarrow\})=\Gamma.
\ee
Finally, we require the pair production rate of defects to be $h$, leading to the condition
\be\label{rate3}
w_j(\{\uparrow\uparrow\uparrow\})=w_j(\{\downarrow\downarrow\downarrow\})=\Gamma h.
\ee
Therefore, our classical stochastic model for the diffusive annihilation process on the 1D lattice is clearly exactly by Eq.~\eqref{master} to Eq.~\eqref{rate2}.

Although we will consider analytical analysis of the master equation later, we can first simulate this effective dynamics numerically with the classical Monte Carlo simulation.   Starting from a random spin configuration $\{\sigma\}$, we randomly update the spin configuration with the three types of random processes (defect hopping, pair annihilation, and pair production) with the corresponding rates given by Eq.~\eqref{rate1} to Eq.~\eqref{rate3}.  By averaging multiple stochastic histories, we can evaluate the physical variables, such as the average defect density $m$ as shown in Fig.~\ref{fig:figure_3} in the main text (where we have chosen $h=0$, $L=100$, several values of $\Gamma$ and averaged over $1000$ histories).  Through fitting, we get the asymptotic power-law decay of the defect numbers $m(t, h=0)$ $\sim (\Gamma t)^{-1/2}$.

 Now we consider the analytical analysis, following the treatment in Ref.~\cite{Racz:1985fy} and \cite{Glauber:1963ij}.  We first propose an approximated form of the transition rates
\be\label{approx}
w_j(\sigma_j)=\frac{1}{2}\Gamma\left[1-\frac{1}{2}\gamma\sigma_j(\sigma_{j+1}+\sigma_{j-1})\right],
\ee 
where we choose $\gamma = (1-h)/(1+h)$.
With this formula, we get the following conditions for the  rates: (1) $w_j(\{\uparrow\uparrow\downarrow\})=w_j(\{\downarrow\downarrow\uparrow\})=w_j(\{\downarrow\uparrow\uparrow\})=w_j(\{\uparrow\downarrow\downarrow\})=\frac{1}{2}\Gamma$, (2) $w_j(\{\uparrow\downarrow\uparrow\})=w_j(\{\downarrow\uparrow\downarrow\})=\frac{1}{2}\Gamma(1+\gamma)= \Gamma/(1+h)$, and (3) $
w_j(\{\uparrow\uparrow\uparrow\})=w_j(\{\downarrow\downarrow\downarrow\})=\frac{1}{2}\Gamma (1-\gamma)=\Gamma h/(1+h)$.  Note that condition (1) here is always the same as Eq.~\eqref{rate1} in our  exact classical stochastic model.   On the other hand, in the $h=0$ ($\gamma=1$) case, conditions (2) and (3) are also the same as Eq.~\eqref{rate2} and ~\eqref{rate3} in the exact classical stochastic model.  For small $h$, condition (2) and (3) remain a good approximation to the exact conditions Eq.~\eqref{rate2} and ~\eqref{rate3}. One can think of this approximation as introducing additional effective short-range repulsion between the defects which slightly decreases the defect annihilation rate.  This short-range repulsion is expected to not change the scaling property in the $h\rightarrow 0$ limit. 

The reason to choose this approximated form in Eq.~\eqref{approx} is due to its equivalence to a kinetic Ising model as first pointed out by Glauber \cite{Glauber:1963ij}.  The Hamiltonian of the Ising model is
\be
H=-J\sum_j \sigma_j \sigma_{j+1}.
\ee
When reaching equilibrium at temperature $T$, the ratio between the reduced probability distributions of the $j$th spin is determined by the Boltzman distribution, i.e.,
\begin{align}\label{ratio1}
\non \frac{p_j(-\sigma_j)}{p_j(\sigma_j)}=&\frac{\exp[(-J/k_BT)\sigma_j(\sigma_{j-1}+\sigma_{j+1})]}{\exp[(J/k_BT)\sigma_j(\sigma_{j-1}+\sigma_{j+1})]} \\
=&\frac{1-\frac{1}{2}\sigma_j(\sigma_{j-1}+\sigma_{j+1})\tanh(\frac{2J}{k_B T})}{1+\frac{1}{2}\sigma_j(\sigma_{j-1}+\sigma_{j+1})\tanh(\frac{2J}{k_B T})}.
\end{align}
On the other hand, the approximated stochastic model defined by Eq.~\eqref{master} and Eq.~\eqref{approx} will approach a steady-state distribution with the ratio
\begin{align}\label{ratio2}
\non \frac{p_j(-\sigma_j)}{p_j(\sigma_j)}=&\frac{w_j(\{\sigma_{j-1},\sigma_j,\sigma_{j+1}\})}{w_j(\{\sigma_{j-1},-\sigma_j,\sigma_{j+1}\})}\\
=&\frac{1-\frac{1}{2}\gamma\sigma_j(\sigma_{j-1}+\sigma_{j+1})}{1+\frac{1}{2}\gamma\sigma_j(\sigma_{j-1}+\sigma_{j+1})}.
\end{align}
By comparing Eq.~\eqref{rate2} with Eq.~\eqref{rate1}, we see that they are exactly the same with the correspondence 
\be\label{effectiveT}
\gamma=\tanh(\frac{2J}{k_B T}).
\ee 
This suggests that the approximated stochastic model is equivalent to a kinetic Ising model, whose steady-state distribution is the same as the equilibrium distribution at the  effective temperature $T$ according to Eq.~\eqref{effectiveT}.  

With the above equivalence, we can get steady state distribution for our exact stochastic model in the small $h$ limit via the equilibrium distribution.  Due to the mapping $m_j$$=$$(1$$-$$\sigma_j \sigma_{j+1})/2$, we can first evaluate the equilibrium distribution of the correlation function in the Ising model, which has the exact solution
\be
\langle \sigma_j \sigma_{j+1} \rangle_\text{eq}=\tanh(J/k_B t).
\ee
Therefore, in the small $h$ limit, we get the average defect density at steady state as
\be
m_s = (1-\langle \sigma_j \sigma_{j+1} \rangle_\text{eq})/2 =\frac{h^{1/2}}{1+h^{1/2}}\approx h^{1/2}.
\ee
Since the correlation $\sigma_j \sigma_{j+1}$ is proportional to energy, its relaxation time $\tau$ can be determined by the long-time decay of homogenous energy perturbation as studied in Ref.~\cite{Felderhof:1971tj}, leading to the result
\be
\tau^{-1}=2\Gamma(1-\gamma)=4\Gamma h/(1+h)\approx 4\Gamma h.
\ee

As we have mentioned in the main text,  in general we have the following three asymptotic scaling laws for the diffusive annihilation process: $m(t, h=0)$$\sim$$(\Gamma t)^{-\xi}$, $m_s \sim h^{1/\delta}$ and $\tau \sim \Gamma^{-1} h^{-\Delta}$. In addition, there is a scaling relation $\Delta\xi=1/\delta$, derived by Ref.~\cite{RACZ:1985kt}. From the property of equilibrium and kinetic Ising models, we have just obtained $\delta =2$ and $\Delta=1$. Using the scaling relation, we can derive that $\xi =1/2$, which has been confirmed by our classical Monte Carlo simulation discussed above and shown in Fig.~\ref{fig:figure_3} in the main text. We also note that exact analytical derivation of $\xi$ can be found in Ref.~\cite{Torney:1983cy}.  

We also note that in the presence of noise inducing pair production of defects ($h>0$), the steady state of the system is a mixed state of different defect configurations, of which the density matrix in the small-$h$ limit can be described by the equilibrium density matrix of the effective Ising model at the corresponding effective temperature $T$ according to Eq.~\eqref{effectiveT}.  The detailed pair condensate description of each defect configuration is encoded by the wavefunctions in Eq.~\eqref{defectwavefunction} in the main text.

So far, we have focused in the context of 1D chain, while we note such a stochastic model and the corresponding master equations can be straightforwardly generalized to higher dimension, as discussed in Ref.~\cite{Lee:1994ic}.  We note for dimension equal or larger than the critical dimension, i.e., $d \ge d_c=2$ \cite{Lee:1994ic}, a simple mean-field description for the diffusive annihilation process gives the correct scaling $m(t, h$$=$$0)$$\sim$$(\Gamma t)^{-1}$  as we have discussed in the main text. 

\section{Derivation for the Circuit-QED Realization}\label{append:circuit}
\begin{figure}[h]
 \includegraphics[width=0.5\textwidth]{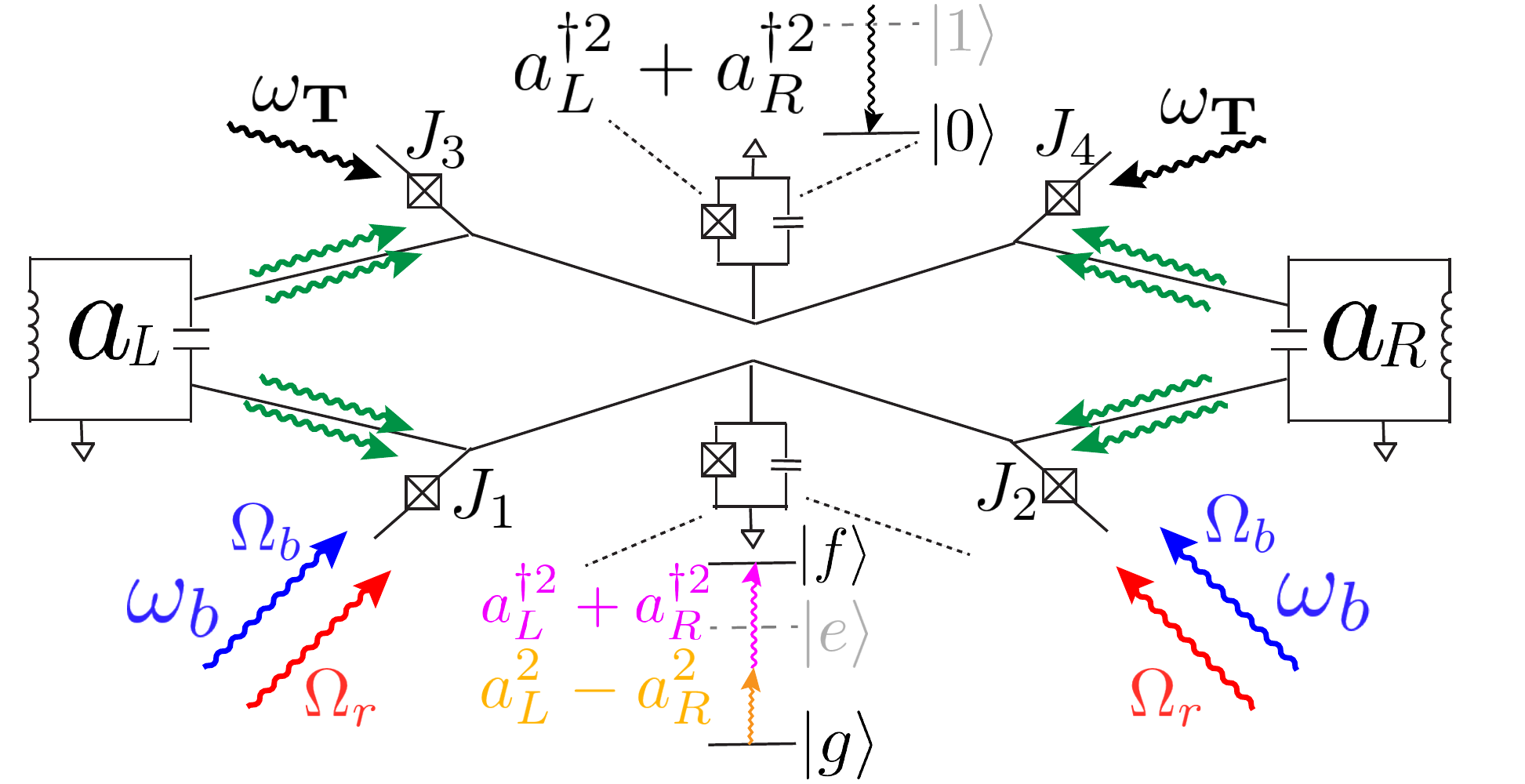}
 \caption{The experimental setup. The junction modes $J_3$, $J_4$ and the two-level system $\ket{0}$ and $\ket{1}$ are introduced to cancel the Kerr nonlinearity induced by $J_1$ and $J_2$. The rest of the components remain the same as we have described in the main text. }
 \label{System_full}
 \end{figure}

The experimental setup is shown in Fig. \ref{System_full}. The two-level system and the two junction modes $J_3$ and $J_4$ are added to cancel the Kerr nonlinearity in the cavity $a_L$ and $a_R$. 

We start by writing the main results of this derivation. The effective dynamics of the system is 
\begin{align}
& \frac{d}{dt}\rho = -i[H,\rho] + \mathcal{D}[l]\{\rho\}\notag\\
&H =  -\chi\sum_{\lambda = L,R}a_\lambda ^{\dagger 2}a_\lambda ^2 + \frac{|g_1|^2}{\delta_1}(a_L^{\dagger 2}-a_R^{\dagger 2})(a_L^2-a_R^2)  \notag \\
&+ \frac{|g_3|^2}{\delta_2}(a_L^{\dagger 2}+a_R^{\dagger 2})(a_L^{2}+a_R^{2})  \notag \\
&l = \sqrt{\frac{2}{\kappa_2}}\frac{g_1 g_2}{\delta_1} (a_L^{\dagger 2}+a_R^{\dagger 2})(a_L^{2}-a_R^{2}).
\label{result_master_eq}
 \end{align}
The coupling constant $g_1$, $g_2$ and $g_3$ will be defined later. All the non-linear terms in the Hamiltonian can be canceled with each others by choosing $\chi = \frac{|g_1|^2}{2\delta_1} = \frac{|g_3|^2}{2\delta_2}$. We can achieve this by tuning the driving frequency $\omega_r$,$\omega_b$ and $\omega_T$. Therefore we can obtain a pure dissipative process that generate photon pair condensate as its steady state.

The Hamiltonian of the two cavities, anharmonic oscillator and the Josephson junction modes. 
\begin{align}
&H = H_0 + V_J + H_d, \notag \\
&H_0 = \omega_c\sum_{\lambda = L,R}a_\lambda^\dagger a_\lambda + \sum_{i=1}^4 \omega_{J}J^\dg_{i} J_{ i} + \sum_{j = g,e,f}\omega_j \ket{j}\bra{j} + \omega_{TLS} \sigma_z , \notag \\
&V_J = - E_{J}\sum_{i = 1}^4 (\frac{\varphi_{i} ^2}{2} + \cos \varphi_{ i}), \notag \\
&H_d = \sum_{i = 1}^4 \Omega_{i}(t) J_{i}^\dagger+h.c.,
\end{align}
where $\omega_c$ is the resonance frequency of the cavity, $\omega_j$ is the energy of anharmonic oscillator which is modeled by the three-level system ($\ket{g}$, $\ket{e}$ and $\ket{f}$), $\omega_{TLS}$ is the resonance frequency of the two-level system and $\omega_{J}$ is the resonance frequency of the Josephson junction modes and we assume that these four Josephson junction modes have the same frequency for simplicity. 

The phases cross the $i$th junction modes are 
\begin{align}
&\varphi_1 = \varphi_c a_L + \varphi_{an}\Sigma + \varphi_{J}J_1  + h.c., \notag \\
&\varphi_2= \varphi_c a_R + \varphi_{an}\Sigma + \varphi_{J}J_2  + h.c., \notag \\
&\varphi_3= \varphi_c a_L + \varphi_{TLS}\sigma + \varphi_{J}J_3  + h.c., \notag \\
&\varphi_4= \varphi_c a_R + \varphi_{TLS}\sigma + \varphi_{J}J_4  + h.c.,
\end{align}
 where $\varphi_a$, $\varphi_{an}$ and $\varphi_{J}$ are the amplitude participation ratios of the respective modes in the junction and $\Sigma$ is the annihilation operator of the anharmonic oscillator and we only consider the lowest three levels. Therefore, $\Sigma = \ket{1}\bra{0} + \sqrt{2} \ket{2}\bra{1}$. The driving fields are defined as
\begin{align}
\Omega_i(t) = 
\begin{cases}
\Omega_{ri} e^{i\omega_rt} + \Omega_{bi} e^{i\omega_bt} , & i = 1,2 \notag \\
\Omega_{Ti} e^{i\omega_Tt}, & i = 3, 4
\end{cases}
.
\end{align}

\subsection{Displaced Frame}
In this section ,we follow the derivation in [1] to move the contribution of the drivings to the phase degree of freedom, we perform the following transformations:
\begin{itemize}
\item 1. Go into the rotating frame of tone $\omega_{r}$. The junction modes becomes $J_{1(2)} \rightarrow J_{1(2)} e^{-i\omega_{r} t}$. 
\item 2. Displace the two junction modes by $J_{1(2)} \rightarrow J_{1(2)} - \frac{\Omega_{r, 1(2)}}{\omega_{J} - \omega_r}$
\item 3. Move back to the original frame. $J_{1(2)} \rightarrow J_{1(2)} e^{i\omega_{r} t}$.
\end{itemize}
After the transformation, the contribution of driving $\omega_r$ is completely absorbed by the phase $\varphi'_{i}(t) = \varphi_\lambda a_\lambda + \varphi_{an}\Sigma_{an} + \varphi_{J}(J_i -  \frac{\Omega_{\lambda, 1}e^{i\omega_{d1}t}}{\omega_c - \omega_{d1}} )  + h.c. $ for i = 1,2. . We can perform the same transformation for driving $\omega_b$ and $\omega_T$ and expand the cosine term to the fourth order. The resulting Hamiltonian becomes
\begin{align}
&H = H_0  - \sum_{i = 1}^4 \frac{E_J}{24} [\tilde{\varphi}_{\lambda}(t)]^4,
\end{align}
where 
\begin{align}
 \tilde{\varphi}_{i}(t) = \varphi_c a_\lambda + \varphi_{an}\Sigma + \varphi_J(J + \xi_i) + h.c. 
 \end{align} 
and
\begin{align}
\xi_{i} = 
\begin{cases}
 \frac{\Omega_{r}}{\omega_{J_i} - \omega_{r}} e^{i \omega_{r} t } + \frac{\Omega_{r}}{\omega_{J_i} - \omega_{b}} e^{ i \omega_{b} t }, & i = 1,2\notag \\
\frac{\Omega_{T}}{\omega_{J_i} - \omega_{T}} e^{ i \omega_{T} t } & i = 3,4
\end{cases}
.
\end{align}

\subsection{Expansion of $ [\tilde{\varphi}_{\lambda} (t)]^4$ }
We now expand the $ [\tilde{\varphi}_{\lambda} (t)]^4$ in order to obtain the desired interaction in the non-rotating frame. After expansion, we perform normal-ordering to the expansion terms, and take all Stark shifts and Lamb shifts into account $\omega \rightarrow \tilde{\omega}$.

By tuning the driving 
\begin{align}
&\omega_{r} = 2\tilde{\omega}_c-\tilde{\omega}_{f}+\tilde{\omega}_{g} - \delta_1, \notag\\
&\omega_{b} = 2\tilde{\omega}_c+\tilde{\omega}_{e}-\tilde{\omega}_{f}  + \delta_1, \notag\\
&\omega_{T} = 2\tilde{\omega}_c -\omega_{TLS} + \delta_2, \notag\\
&\Omega_{r,1} = -\Omega_{r,1}, \notag \\
&\Omega_{b,2} = \Omega_{b,2}, \notag \\
&\Omega_{T3} = \Omega_{T,4}
\end{align}
and performing rotating wave approximation, we have
\begin{align}
H &= V_{kerr} + \delta_1 \ketbra{e} + \delta_2 \sigma_{z}+ [g_1(a_L^{2}-a_R^{2})\ket{e}\bra{g} \notag \\
&+ g_2(a_L^{\dagger 2}+a_R^{\dagger 2})\ket{f}\bra{e} + g_3(a_L^{2}+a_R^{2})\ket{1}\bra{0} + h.c. ] ,
\end{align}
where 
\begin{align}
V_{kerr} = -\chi\sum_{\lambda = L,R}a_\lambda ^{\dagger 2}a_\lambda ^{ 2} -  \chi_{a, an}(a_L^\dagger a_L + a_R^\dagger  a_R)(\ket{e}\bra{e} + 2\ket{f}\bra{f}) \notag \\
- \chi_{a, T}(a_L^\dagger a_L + a_R^\dagger  a_R)\ket{1}\bra{1},
\end{align}
where the Kerr-nonlinearities are $\chi = E_J\varphi_c^4$, $\chi_{a,an} = E_J\varphi_c^2 \varphi_{an}^2$ and $\chi_{a,T} = E_J\varphi_c^2 \varphi_{T}^2$ .

\subsection{Large Detuning Limit }
Next, when $\delta_1, \delta_2 \gg g_1, g_2, \chi$, we can adiabatically eliminate the level $\ket{f}$ and $\ket{1}$ by performing the Schrieffer-Wolff transformation. The unitary operator is
\begin{align}
U &= \exp(-S), \notag \\
S &= -\frac{g_1}{\delta_1}(a_L^{2}-a_R^{2})\ket{e}\bra{g} + \frac{g_2}{\delta_1}(a_L^{\dagger 2}+a_R^{\dagger 2})\ket{f}\bra{e}, \notag \\
&-\frac{g_3}{\delta_2}(a_L^{2}+a_R^{2})\ket{1}\bra{0} - h.c. .
\end{align}
By keeping the terms up to $O(\frac{g_{1(2)}^2}{\delta})$ and neglecting the terms proportional to $\ketbra{e}$ and $\ketbra{1}$, we have
\begin{align}
H' &= e^{-S}He^S\notag \\ 
&\approx 
V'_{kerr}  + \frac{|g_2|^2}{\delta_1}(a_L^{\dagger 2}+a_R^{\dagger 2})(a_L^{2}+a_R^{2})\ket{f}\bra{f}\notag \\ 
&+ \frac{|g_1|^2}{\delta_1}(a_L^{\dagger 2}-a_R^{\dagger 2})(a_L^{2}-a_R^{2})\ket{g}\bra{g} \notag \\ 
&+  [\frac{g_1 g_2}{\delta_1} (a_L^{\dagger 2}+a_R^{\dagger 2})(a_L^{2}-a_R^{2})\ket{f}\bra{g} + h.c. ],
\end{align}
where
\begin{align}
V'_{kerr}  = -\chi(a_L^{\dagger 2}a_L^{ 2} + a_R^{\dagger 2}a_R^{2})  + 2\chi_{a, an}(a_L^\dagger a_L + a_R^\dagger  a_R) \ketbra{f} \notag \\
+ \frac{|g_3|^2}{\delta_2}(a_L^{\dagger 2}+a_R^{\dagger 2})(a_L^{2}+a_R^{2}).
\end{align}

\subsection{Large $\kappa_f$ Limit }
Then we take the dissipation process of the anharmonic oscillator $\ket{f}\rightarrow \ket{g}$ and $\ket{1}\rightarrow \ket{0}$ into account. The dynamics of the system can be described by the master equation
\begin{align}
\frac{d}{dt}\rho = -i[H',\rho] +\kappa_{f} \mathcal{D}[\ket{g}\bra{f}]\{\rho \},
\end{align}
where $\kappa_{f}$ is the decay rate from state $\ket{f}$ to $\ket{g}$ . Assume that the decay rate $\kappa_{f} \gg \frac{|g_1|^2}{\delta_1}, \frac{|g_2|^2}{\delta_1}, \frac{|g_1 g_2|}{\delta_1},  \frac{|g_3|^2}{\delta_2}$, we can derive the effective dynamics of the two cavities.

Consider an operator $\hat{O}$ that only depends on the degrees of freedom of cavities. The equation of motion of $\hat{O}$ can be written as

\begin{align}
\frac{d}{dt}\hat{O} &= -i([\hat{O}, \hat{P}]\ket{f}\bra{f} + [\hat{O}, \hat{Q}]\ket{g}\bra{g} 
\notag \\
&+ [\hat{O}, \hat{R}]\ket{f}\bra{g} + [\hat{O}, \hat{R}^\dagger]\ket{g}\bra{f} + [\hat{O}, V'_{kerr}]),
\label{EOM_O}
\end{align}
where 
\begin{align}
\hat{P} = \frac{|g_2|^2}{\delta}(a_L^{\dagger 2}+a_R^{\dagger 2})(a_L^{2}+a_R^{2}), \notag \\
\hat{Q} = \frac{|g_1|^2}{\delta}(a_L^{\dagger 2}-a_R^{\dagger 2})(a_L^{2}-a_R^{2}), \notag \\
\hat{R} = \frac{g_1 g_2}{\delta} (a_L^{\dagger 2}+a_R^{\dagger 2})(a_L^{2}-a_R^{2}).
\end{align}

Also, the dynamics of $\ket{g}\bra{f}$ is
\begin{align}
\frac{d}{dt}\ket{g}\bra{f} &= -\kappa_f \ket{g}\bra{f} \notag \\
& -i[\hat{P}\ket{g}\bra{f} - \hat{Q}\ket{g}\bra{f} + \hat{R}(\ket{g}\bra{g}-\ket{f}\bra{f})].
\end{align}
Since $\kappa_f$ is much greater than all the other parameters in the effective Hamiltonian, we can assume that $\langle \ket{g}\bra{g} \rangle \approx 1$, $\langle \ket{f}\bra{f} \rangle \ll 1$ and $\langle \ket{f}\bra{g} \rangle \approx O(\frac{g^2}{ \kappa_f \delta})$. In this regime, the operator $\ket{g}\bra{f}$ reaches its stationary state in the time scale that is much smaller than the dynamics of the cavity operator $\hat{O}$. Thus, we can replace the operator $\ket{g}\bra{f}$ by its stationary state
\begin{align}
\ket{g}\bra{f} \approx \frac{-2iR}{\kappa_f}.
\end{align}
Therefore, Eq.(\ref{EOM_O}) becomes
\begin{widetext}
\begin{align}
\frac{d}{dt}\hat{O} &= -i[\hat{O},  \frac{|g_1|^2}{\delta}(a_L^{\dagger 2}-a_R^{\dagger 2})(a_L^{2}-a_R^{2})] -i[\hat{O}, \frac{\chi}{2}(a_L^{\dagger 2}a_L^{ 2} + a_R^{\dagger 2}a_R^{ 2})] \notag \\ 
&- \frac{2}{\kappa_2}(\frac{g_1 g_2}{\delta})^2[\hat{O}, (a_L^{\dagger 2}+a_R^{\dagger 2})(a_L^{2}-a_R^{2})]\times(a_L^{\dagger 2}-a_R^{\dagger 2})(a_L^{2}+a_R^{2}) + h.c..\notag\\
\end{align}
\end{widetext}
The equation of motion can be converted to the master equation of the cavity mode as described in Eq.\eqref{result_master_eq}.

\end{appendix}

\bibliography{mybib.bib}

\end{document}